11

# Cryptic Genetic Variation Can Make Irreducible Complexity

# a Common Mode of Adaptation

Authors
**Authors:**  M.V. Trotter[1,3], D.B. Weissman[2], G.I. Peterson[3], K.M. Peck[3,4], J. Masel[3*]

**Author Affiliations:**

[1]Department of Biology, Stanford University, Stanford, CA, USA 95306.

[2]Institute of Science and Technology Austria, Klosterneuburg, Austria

[3]Department of Ecology and Evolutionary Biology, University of Arizona, Tucson, AZ, USA 85721.

[4]Department of Biology, University of North Carolina at Chapel Hill, Chapel Hill, NC, USA 27599.

***Corresponding author**: masel@u.arizona.edu,  520 626 9888

**Other emails** (in order): mtrotter@stanford.edu, dbw@ist.ac.at, gpeterson81@gmail.com, kaylap@live.unc.edu

**Running title:** Complex adaptation via crypticity

**Keywords:** Complex adaptation, adaptive valley, evolutionary capacitance

**Words**: 4,500

**Figures**: 3

**Supporting Information**: 5 Figures

**Abstract**

The existence of complex (multiple-step) genetic adaptations that are 'irreducible' (i.e., all partial combinations are less fit than the original genotype) is one of the longest standing problems in evolutionary biology. In standard genetics parlance, these adaptations require the crossing of a wide adaptive valley of deleterious intermediate stages. Here we demonstrate, using a simple model, that evolution can cross wide valleys to produce 'irreducibly complex' adaptations by making use of previously cryptic mutations. When revealed by an evolutionary capacitor, previously cryptic mutants have higher initial frequencies than do new mutations, bringing them closer to a valley-crossing saddle in allele frequency space. Moreover, simple combinatorics imply an enormous number of candidate combinations exist within available cryptic genetic variation. We model the dynamics of crossing of a wide adaptive valley after a capacitance event using both numerical simulations and analytical approximations. Although individual valley crossing events become less likely as valleys widen, by taking the combinatorics of genotype space into account, we see that revealing cryptic variation can cause the frequent evolution of complex adaptations. This finding also effectively dismantles 'irreducible complexity' as an argument against evolution by providing a general mechanism for crossing wide adaptive valleys.

# Introduction

When a population is well adapted to its environment, the vast majority of new mutations will be neutral or negative. If a higher fitness genotype exists that requires multiple mutations, but each intermediate mutation combination is deleterious, the population must traverse a metaphorical "adaptive valley" of low fitness to access the superior adaptation (Wright 1932). Such adaptations are called "irreducibly complex" by the intelligent design lobby, which uses the term to assert that evolution cannot cross multi-step adaptive valleys. Detailed investigations into the evolution of specific complex adaptations (Bridgham et al. 2006; Weinreich et al. 2006; Poelwijk FJ et al. 2007; Egelman 2010) have shown that in these particular cases, evolved complexity is not "irreducible". Many biologists assume, in agreement with the intelligent design lobby, that irreducible complexity rarely, if ever, evolves.

In fact, valley crossing in asexual populations is both possible and relatively well understood (Van Nimwegen and Crutchfield 2000; Weissman et al. 2009). In small populations, individually deleterious mutations may fix sequentially by drift (Wright 1932). In large populations, fit multiple-mutants occasionally appear even when the component mutations are rare. This process is called 'stochastic tunneling' (Carter and Wagner 2002; Komarova et al. 2003; Iwasa et al. 2004; Weinreich and Chao 2005; Burton and Travis 2008). Weakly deleterious mutations may also act as stepping stones across deeper adaptive valleys (Covert et al. 2013).

The evolution of complex adaptations is more problematic in sexual populations because of genetic recombination. While recombination can facilitate complex adaptation by bringing together mutations from different lineages into a single individual (Fisher 1930; Muller 1932) it also breaks up beneficial combinations, rendering the crossing of even narrow valleys impossible (Crow and Kimura 1965; Eshel and Feldman 1970; Karlin and McGregor 1971). At low frequencies, mutations required for a given complex adaptation are almost always present separately, where selection acts against them. Rare individuals carrying a complex adaptation are unlikely to mate with other such (rare) individuals, and so produce maladapted offspring. In large populations the situation is particularly dire, as mutations are kept even rarer by more efficient selection. Thus, barring tiny effective population sizes or large mutation rates, high rates of recombination prevent valley crossing (Weissman et al. 2010). This raises the question: is it possible for sexual populations to produce irreducibly complex adaptations at all?

One mechanism that may allow the evolution of complex adaptations is the revelation of cryptic variation (Phillips 1996; Hansen et al. 2000; Masel 2006; Kim 2007), via a phenomenon known as evolutionary capacitance (Bergman and Siegal 2003; Masel 2005; Schlichting 2008; Masel and Trotter 2010). When the environment changes and organisms are stressed, evolutionary capacitors switch the status of genetic variation from "off" (phenotypically cryptic) to "on". After revelation by a capacitor, this previously phenotypically silent genetic variation can acquire fitness consequences, producing a burst of "new" genotypic effects that are potentially adaptive in the new environment. A growing body of both theoretical and laboratory work suggests that such revelation events are a common feature of biological systems (Bergman and Siegal 2003; Jarosz and Lindquist 2010; Tirosh et al. 2010; Freddolino et al. 2012; Janssens et al. 2013; Takahashi 2013).

The argument that crypticity can facilitate complex adaptation is twofold. First, cryptic mutations attain higher allele frequencies than they would if selection against them was operating at full strength. Since allele frequency must exceed a threshold before recombination produces peak genotypes more often than it breaks them up (Weinreich and Chao 2005), high initial frequencies of cryptic mutants can give adaptation a head start across a valley (Kim 2007).

Second, revelation of cryptic variants allows the population to sample genotypes from many new and different parts of genotype space simultaneously. The new genotypes will mostly fall into low fitness valleys, but may, on rare occasions, hit upon new adaptive peaks. As the number of newly exposed mutant loci increases, the number of ways to combine those loci to form potential complex adaptations can become enormous. In other words, while any given complex adaptation is unlikely to fix, simple combinatorics imply that an enormous number of candidate combinations exist within available cryptic genetic variation, any one of which might fix. As well as increasing the (still low) likelihood of any given valley crossing event, crypticity multiplies the number of possible valley crossing events dramatically. In short, many more complex than simple possibilities exist in genotype space. This combinatoric argument has occasionally been made verbally (Fisher 2007; Weissman et al. 2009) but has never been formalized.

Here we show, using a simple population genetic model, that irreducibly complex adaptations can arise and fix under biologically reasonable conditions. We model crossing a fitness valley (Figure 1) in two stages, corresponding to 'before' and 'after' an event in which environmental change prompts revelation via evolutionary capacitance. First, we sample initial allele frequencies at $j$ cryptic mutant sites, using distributions calculated using the Moran model (Masel 2006). Second, we expose those $j$ loci to stronger selection, as expected after revelation by a capacitor. Assuming that the mutations are individually deleterious, but in full combination confer an adaptive advantage in the new environment, we evaluate whether or not the population fixes all $j$ mutant alleles. Simulating many such valley crossing attempts, and counting the successes, yields the probability of crossing a given $j$-step valley after a capacitance event. Each valley is defined by its "width" (number of mutant sites, $j$), its "depth" (the strength of selection against individual revealed mutations, $s_{valley}$), and its "height" (the strength of selection in favor of the new peak genotype, $s_{peak}$).

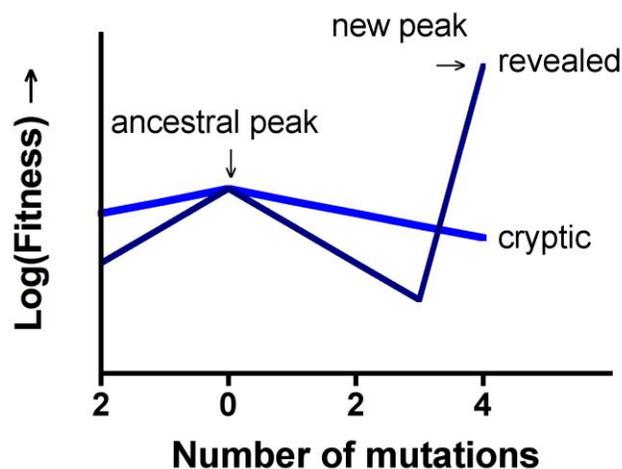

**Figure 1.** A schematic example of a 4-step fitness valley before and after revelation of cryptic variation.

For simplicity, we neglect any fitness effects at other loci. Each additional mutant site incurs a fitness decrement regardless of crypticity, but this decrement is smaller when a mutation is in a cryptic state. Crypticity effectively flattens the valley, allowing the population to spread further across genotype space than would be possible under full-strength selection.

The probability of valley crossing depends on the parameters ($j$, $s_{valley}$, $s_{peak}$) associated with the valley itself, as well as population size ($N$), mutation rate ($\mu$), and selection against new mutations while they are cryptic ($s_{cryptic\_valley}$). Finally, we multiply the probability of crossing a given $j$-step valley by the expected number of available $j$-step genotypes (as derived in (Masel 2006), each of which might be a peak with low probability $\varepsilon$.

Although individual valley crossing events become less probable as valleys widen, by taking the combinatorics of genotype space into account, we find that revealing cryptic variation can cause the frequent evolution of complex adaptations. We also present analytical approximations that agree with our simulation results, providing additional support for our findings.

## Methods

Our model follows a population of $N$ diploid, randomly mating individuals with discrete generations. We assume infinite sites with total mutation rate $\mu$, and free recombination. Fitness effects are multiplicative between sites, and the dominance coefficient is $h=0.5$. Let selection against mutant alleles while in the cryptic state be $s_{cryptic\_valley}$, and while revealed, $s_{valley}$. Selection for adaptive peak genotypes is $s_{peak}$. Where required for brevity, we will write these as $s_{cv}$, $s_v$, and $s_p$. At cryptic sites, we assume weakened selection such that $s_{valley} > s_{cryptic\_valley} > 1/2N$. We use both numerical simulations and analytical approximations to estimate the expected number of complex fixations produced after the revelation of variation by a capacitor, under a wide range of valley and population parameters.

**Simulations**

*Initial "Revealed" Allele Frequencies*

For a given population size, allele frequencies at each of the $j$ sites at the time of revelation were sampled from the distribution of expected sojourn times, with $P(\text{frequency} = i) = \frac{\tau_i(N, s_{cv})}{\tau}$. The expected sojourn times $\tau_i$ during which the allele frequency is $i$ and the total time $\tau$ during which it has not yet become fixed or gone extinct, given that it entered the population as a single new mutation, are calculated from the Moran model, as follows. Strictly speaking, the Moran model applies to haploids only. However, the diploid case can be closely approximated by a model of $2N$ haploids. Before revelation, effects at different sites are independent, with dominance coefficient $h = 0.5$: these assumptions are necessary in order to approximate a diploid population using the Moran model. Given that homozygosity of mutant cryptic alleles is extremely rare, relaxing the codominance assumption is not expected to be important.

At each time step before revelation, one haploid individual is randomly chosen to die and one to reproduce. $2N$ such time steps comprise one "generation". For a given polymorphic site at time $t$, $i$ copies of the mutant allele will be present in the population, each contributing factor $1-s_{cv}$ to the fitness of the individual that carries them. The evolutionary dynamics are dominated by the selection coefficient on one copy of a mutant allele, set here to be $s_{cv}$. We model fitness as differential reproduction. As an infinite sites model, it is possible to neglect recurrent and back mutations, so the next reproducing individual has the mutant allele with probability:

$$\frac{(1-s_{cv})i}{2N-i+(1-s_{cv})i}$$

$$= \frac{(1-s_{cv})i}{2N-s_{cv}i}$$

The probability that the mutant allele appears in the next haploid individual chosen to die is $i/2N$. The probability that the number of copies of the mutant allele increases from $i$ by 1 is then given by the probability that a mutant haploid individual reproduces while a wild-type haploid individual dies:

$$\lambda_i = \frac{(1-s_{cv})i(2N-i)}{(2N-s_{cv}i)2N}$$

The probability that the number of mutants decreases from $i$ by 1 then takes into account the probability that a mutant individual dies and a wild-type individual reproduces:

$$\mu_i = \frac{i(2N-i)}{2N(2N-s_{cv}i)}$$

Following Ewens ((Ewens 2004), eqn 2.158), let

1 $$\rho_o = 1, \rho_i = \frac{\prod_{k=1}^{i} \mu_k}{\prod_{k=1}^{i} \lambda_k} = (1-s)^{-i}$$

2 The probability of fixation by drift starting from *i* individuals is then $\pi_i = \frac{\rho_i - 1}{\rho_{2N} - 1}$ and the

3 probability of fixation by drift starting from a single mutant individual is $p_{\text{fix}} = \frac{\rho_1 - 1}{\rho_{2N} - 1}$ ((Ewens

4 2004), eqn 2.159). The sojourn time $\tau_i$ during which there are *i* descendants of a single original

5 mutant is given by

6 $$\tau_i = \frac{\rho_1(\rho_{2N-i} - 1)(2N - s_{cv}i)}{(\rho_{2N} - 1)i(2N - i)}, i = 1, 2, \ldots, 2N - 1$$

7 and $\tau = \sum_{i=1}^{2N-2} \tau_i$, (37, eq 2.144), where the unit of time is one generation or *2N* rounds in the

8 Moran model.

9 For each of *j* mutant sites, allele frequencies were sampled from the sojourn time

10 distributions as above, to be used as the initial frequencies of "revealed" cryptic variants in the

11 second part of our simulations. Alleles were assigned randomly to individuals, such that there

12 was no initial linkage disequilibrium, and genotype frequencies at each site followed

13 approximately Hardy-Weinberg proportions. Mean linkage disequilibrium is generally slightly

14 negative in asexual populations, but of negligible magnitude in sexual populations (Kouyos et al.

15 2007).

17 *Selection after revelation of cryptic variants*

18 We model two different dominance scenarios. First, we assume the mutant alleles to be recessive

19 for their adaptive function, such that only individuals with two mutant alleles at each site acquire

20 the new peak fitness. This is the most conservative case of valley crossing, which we call a

21 recessive peak. In the case of a dominant peak, individuals with at least one mutant allele at each

22 site acquire the fitness peak. Thus, dominant adaptive peaks are slightly more likely to fix.

After revelation, the j-site genotypes were assigned new fitness based on the number of copies of mutant alleles. In the case of a recessive peak:

$$w(\text{valley}) = (1 - s_v)^k, k = \text{total number of mutant alleles across } j \text{ sites}$$

$$w(\text{peak}) = 1 + s_p, \text{ if } k = 2j$$

In the case of a dominant peak:

$$w(\text{valley}) = (1 - s_v)^k, k = \text{total number of mutant alleles across } j \text{ sites}$$

$$w(\text{peak}) = 1 + s_p, \text{ if } k = 2j, \text{ if at least one mutant allele is present at each site.}$$

Note that individuals with no mutant alleles maintain a fitness of 1 both before and after revelation.

Each generation of our model has three steps: recombination/reproduction, viability selection and drift.

In the reproduction step, each genotype contributes gametes to an infinite gamete pool according to its frequency in the population, with free recombination. For a system with $j$ mutant loci, we have $k = 3^j$ multilocus genotypes capable of producing $2^j$ possible gametes. After recombination, new genotype frequencies were generated by random union of gametes. Viability selection was then applied to the new genotypes based on genotype fitnesses, such that for each multisite genotype frequency $g_i, g_{i+1}, \ldots g_k$ the frequency after viability selection is given by:

$$g_i' = \frac{g_i w_i}{\sum_{p=1}^{k} g_p w_p} = \frac{g_i w_i}{\bar{w}}$$

Finally, genotype frequencies were resampled from a multinomial distribution to simulate sampling effects due to finite population size.

Once any of the $j$ mutant sites went extinct, the $j$-site peak became inaccessible and so the simulation was stopped and the population recorded as fixed for the original wild type peak. The population was considered fixed for the new peak if all individuals in the population carried two copies of mutant alleles at all $j$ sites.

For each combination of $s_{cv}$, $s_v$, and $s_p$, the population was iterated to fixation/extinction from at least $10^5$ sets of initial allele frequencies and, in the case of deeper valleys, from up to $10^7$ sets (so as to detect even very rare fixations). This probability is then multiplied by the expected number of newly available $j$-step genotypes following a capacitance event, to arrive at our final estimate of the expected number of complex fixations (E[fixations]).

*Expected number of potential peaks*

The expected number of available polymorphic sites in the population is Poisson distributed around a mean of $2\mu N \tau (N, s_{cryptic\_valley})$ where $2\mu N$ is the rate of introduction of mutant sites, and $\tau$ is the expected sojourn time of a new mutation under the Moran model. The expected number of potential peaks (combinations of $j$ alleles) available to the population is then (Masel 2006):

$$\frac{\left(2\mu N \tau(N, s_{cryptic\ valley})\right)^j}{j} \tag{1}$$

We assume that any one of the many possible combinations of $j$ sites could be adaptive, each with low probability ε. All adaptation rates are proportional to the infinitesimal ε, which can be approximated as constant and hence factor out of their ratios unless ε scales extremely strongly with $j$.

**Analytical Approximations**

We would like to find a mathematical approximation for the probability of irreducibly complex adaptation from formerly cryptic variation. In our model, this probability is the product of two factors: ε (the probability that a given combination of mutations is adaptive), and E[fixations], the expected number of combinations of j mutations that would fix if they were adaptive. We want to find this second factor. We do this by writing it as the probability that at the time that cryptic genetic variation is revealed, there is a set of *j* mutations present at frequencies $x_1 \geq x_2 \geq \ldots x_j$, multiplied by the probability that they successfully fix (assuming that they are an adaptive combination), summed over all possible combinations of $x_i$.

$$E[fixations] = \sum_{x_1 \geq \ldots x_j} \left( P(frequencies\ x_1, \ldots x_j) * P(fixation|x_1, \ldots x_j) \right) \quad (2)$$

Finding the first factor in (2) is straightforward. Let $\phi(x)$ be the equilibrium site-frequency spectrum of the cryptic variation, meaning that the probability that there is a mutation with frequency in the infinitesimal range $[x, x+dx]$ is $\phi(x)dx$, or equivalently, that the expected number of mutations with frequencies between $y_1$ and $y_2$ is $\int_{y_1}^{y_2} \phi(x)dx$. Using the diffusion approximation, $\phi$ is given by (Ewens, 9.23, adjusted for a Moran model):

$$\phi(x) = \frac{2N\mu}{x(1-x)} \frac{e^{2Ns_{cv}x} - e^{2Ns_{cv}}}{1 - e^{Ns_{cv}}} \quad (3)$$

Since all the mutations are independent, the joint spectrum is the product $\prod_{i=1}^{j} \phi(x_i)$. Note that since the diffusion approximation requires continuous allele frequencies, we must also change the sum in (2) to an integral when we substitute in (3).

$$E[fixations] \approx \int_{x_1 \geq \ldots x_j} dx \prod_{i=1}^{j} \phi(x_i) * P(fixation|x_1, \ldots x_j) \quad (4)$$

Calculating the second factor in (2), the probability of fixation given the vector of initial frequencies, is harder. To make progress we first assume that the trajectory of the mutations will be entirely determined by selection once the variation is revealed, so that we can neglect stochastic effects. This will be accurate as long as mutations that start out at very low frequencies do not contribute much to the probability of complex adaptation. With this assumption, the probability of fixation is always 0 or 1, so instead of having to find and multiply it, we can take an integral over the basin of attraction of fixation (which we here call *V*) under deterministic expectations.

Thus, our approximation takes the form:

$$E[\text{fixations}] \approx \int_V dx \prod_{i=1}^{j} \phi(x_i) \qquad (5)$$

However, finding the boundary of this basin of attraction is difficult, and in general must be done approximately. Let $\bar{s}_i$ be the mean advantage of mutation *i* over the wild-type allele. Since $x_1 \geq x_2 \geq \ldots x_j$, it follows $\bar{s}_1 \leq \bar{s}_2 \leq \cdots \bar{s}_j$ (because the advantage of each mutation increases as the frequencies of the other mutations increase). Thus, a necessary condition for being in the basin of attraction of fixation is to have $\bar{s}_j > 0$ (i.e., for the rarest mutant to be initially favored), and a sufficient condition is to have $\bar{s}_1 > 0$ (i.e., for the most common mutant to be favored). We find the expected number of sets of mutations satisfying the latter condition for both recessive and dominant adaptive peaks, keeping in mind that this gives an underestimate of E[fixations].

*Recessive adaptations*

First consider the case in which the complex adaptation is recessive. In this case, the mean selective advantage of mutation *i* is:

$$\bar{s}_i = s_v + 2(s_p + s_v) x_i \prod_{k \neq i} x_k^2$$

If we assume $s_p \gg s_v$, (the peak is more advantageous than the valley is disadvantageous), this is approximately:

$$\bar{s}_i \approx s_v + 2 s_p x_i \prod_{j \neq i} x_j^2$$

The condition for all $\bar{s}_i$ to be positive is therefore

$$x_1 \prod_{i=2}^{j} x_i^2 > \frac{s_v}{2s_p} \qquad (6)$$

Using the region defined by (6) as an approximation for $V$, (2) becomes

$$E[\text{fixations}] = \int_{x_1^{min}}^{1} dx_1 \int_{x_2^{min}}^{x_1} dx_2 \ldots \int_{x_j^{min}}^{x_{j-1}} dx_j \prod_{i=1}^{j} \phi(x_i) \qquad (7)$$

where the lower bounds for the integrals are given by

$$x_1^{min} = \left(\frac{s_v}{2s_p}\right)^{\frac{1}{2(j-1)}}$$

and

$$x_i^{min} = \left(\frac{s_v x_1}{2s_p \prod_{k=1}^{i-1} x_k^2}\right)^{\frac{1}{2(j-i+1)}} \text{ for } i \geq 2$$

*Dominant adaptations*

If the complex adaptation is dominant, the mean selective advantage of mutation $i$ is

$$\bar{s}_i = s_v + 2^{j-1}(s_p + s_v) x_i \prod_{k \neq i} x_k (1 - x_k)$$

and if we assume $s_p \gg s_v$, and that all mutations are initially at low frequencies

$$\bar{s}_i \approx -s_v + 2^{j-1} s_p \prod_{k \neq i} x_k$$

With these approximations, the condition for all $\bar{s}_i$ to be positive is

$$\prod_{i=2}^{j} x_i > \frac{s_v}{2^{j-1} s_p}$$

The approximate expected number of potential combination is still given by (7), but now the lower bounds on the integral are given by

$$x_1^{min} = \frac{1}{2}\left(\frac{s_v}{s_p}\right)^{\frac{1}{j-1}}$$

and

$$x_i^{\min} = \left(\frac{s_v x_1}{2^{j-1} s_p \prod_{k=1}^{i-1} x_k}\right)^{\frac{1}{j-i+1}} \text{ for } i \geq 2$$

However, using these bounds can substantially underestimate *V*. This is because there is a large contribution from very small values of $x_j$, where our approximations break down. (In contrast, for recessive adaptations, $x_j$ cannot drop too low, because then there would be no individuals homozygous for the j[th] mutation.) To account for finite population size, we must adjust the integral bounds to also include the requirement that $x_j > 1/2N$, to avoid impossibly low allele frequencies.

We also need to have a better approximation for the volume of *V* that includes frequencies such that the most common mutant alleles are initially disfavoured, but then switch to being favoured as the rarer mutant alleles increase in frequency. Rather than requiring that the lowest frequency $x_j$ be high enough for all alleles to have an initial advantage, we can just require that it be within striking distance of this threshold. Specifically, we can require that once variation is revealed, it will increase to the value given by $x_i^{\min}$ before the other $x_i$ decrease much. These other alleles will have a mean fitness disadvantage of at most $s_v$, so we can make the approximation that $x_j$ has about $1/s_v$ generations to reach the threshold before they can decrease significantly. During this time, we can approximate the j[th] allele's advantage by $\bar{s}_i \approx s_v + 2^{j-1} \frac{s_p}{|s_v|} \prod_{k<j} x_k$, with the $x_k$ taken to be their initial values. The adjusted initial minimum value for $x_j$ is now

$$\hat{x}_i^{\min} \approx \frac{|s_v| x_1}{2^{j-1} s_p \prod_{k<j} x_k} \exp\left[-1 + 2^{j-1} \frac{s_p}{|s_v|} \prod_{k<j} x_k\right] \qquad (8)$$

1   Using these approximations, numerical integration of (7) gives the lower $j = 3$ curve in Figure 2.

2   For $j \geq 4$, however, even more accurate approximations are needed, and this approach becomes

3   impractical.

4   For the simplest case, $j = 2$, we can do better, and actually find an "exact" expression for

5   the region in which the mutations will be driven to fixation:

6   $$x_2 > x_1 \exp\left(-\frac{2s_p(x_1 - x_2)}{s_v}\right), \text{ and } x_1 \geq x_2$$

7   Integrating $\phi(x_1)\phi(x_2)$ over this region gives the lower $j = 2$ curve in Figure 2.

9                                          **Results**

10  We find that, for a large subset of parameter value combinations, the expected number of

11  fixation events (Figure 2) is much greater for complex adaptations than for simple one step

12  adaptations, even given a recessive peak. In both cases illustrated in Figure 2, as crypticity

13  becomes stronger (small $s_{cryptic\_valley}$), a higher proportion of adaptations are complex. Higher

14  values of $j$ represent valleys that are more difficult to cross, but they also present the population

15  with a larger number of possible peaks to sample. For strong crypticity, the latter effect

16  dominates the results, leading to many complex fixation events.

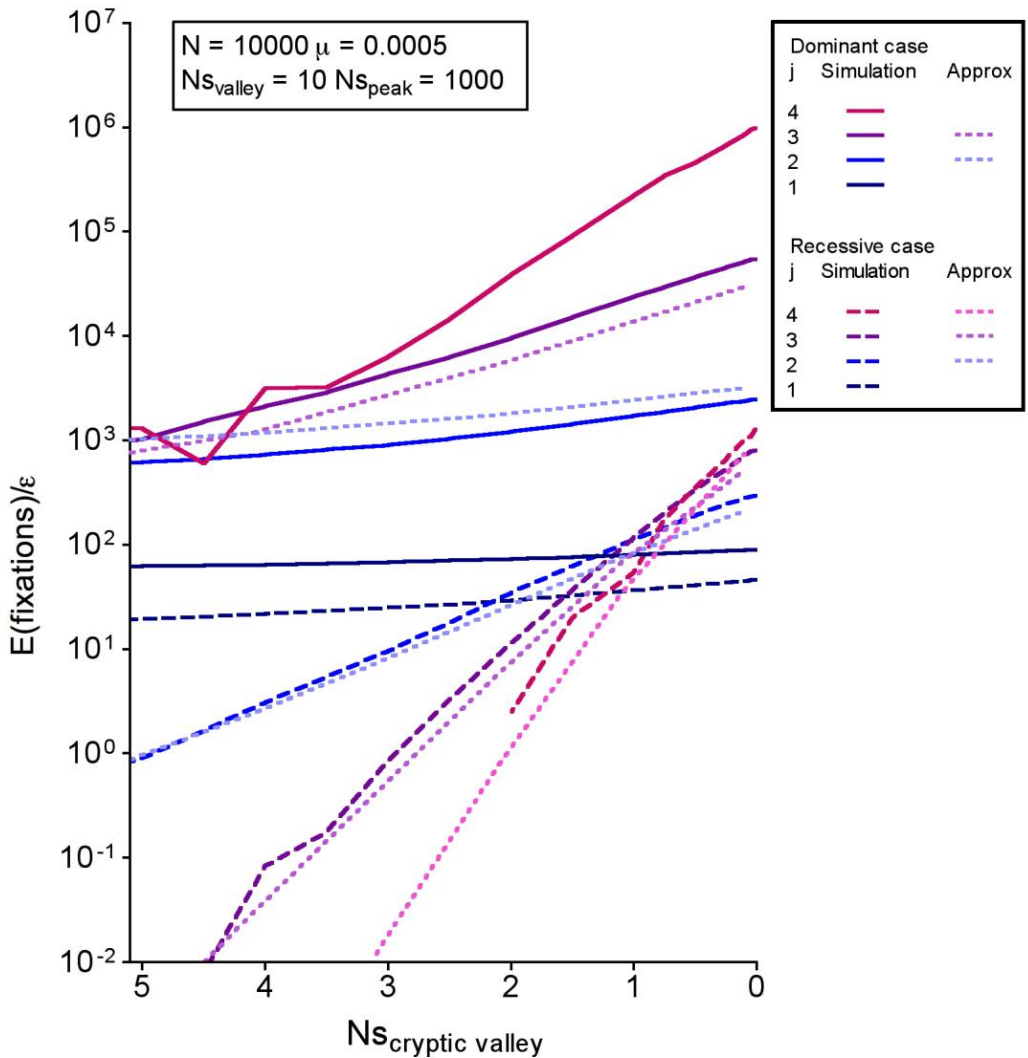

**Figure 2.** Expected numbers of fixed $j$-sized adaptations as a function of crypticity. Crypticity is the strength of selection on cryptic alleles, and can be interpreted either as penetrance or as expressivity. For each value of $j$ and $s_{cryptic\_valley}$, we performed at least $10^5$ simulations to calculate a Monte Carlo estimate of the probability that a given $j$-sized adaptation would fix, and multiplied this by the expected number of $j$-sized adaptations that could be created by recombination of existing polymorphic alleles. Lack of smoothness represents limited statistical resolution for rarely fixing complex adaptations. Approximations were evaluated using numerical integration in Mathematica.

The number of potential peaks depends on the $j^{th}$ power of the expected number of segregating mutant sites (Eq. 1), which in turn depends on the mutation rate, on the population size, and on the sojourn time. When crypticity is strong ($s_{cryptic\_valley}$ is small), mutations are nearly neutral and sojourn times are dominated by $N$. When $s_{cryptic\_valley}$ is large and crypticity is weak, selection shortens sojourn times. Exponential dependences can lead to an abrupt transition to the complex-adaptation regime, e.g. with a threshold value of $s_{cryptic\_valley} \sim 1/N$ in the recessive case shown in Figure 2. If selection on cryptic variation exceeds this threshold, sojourn times are so short that very few segregating sites exist at any given moment, and thus few potential peaks are available.

Our results are best summarized by looking at the proportion of all expected fixations that have $j > 1$. This proportion, for a range of mutation rates and levels of crypticity, is illustrated in Figure 3.

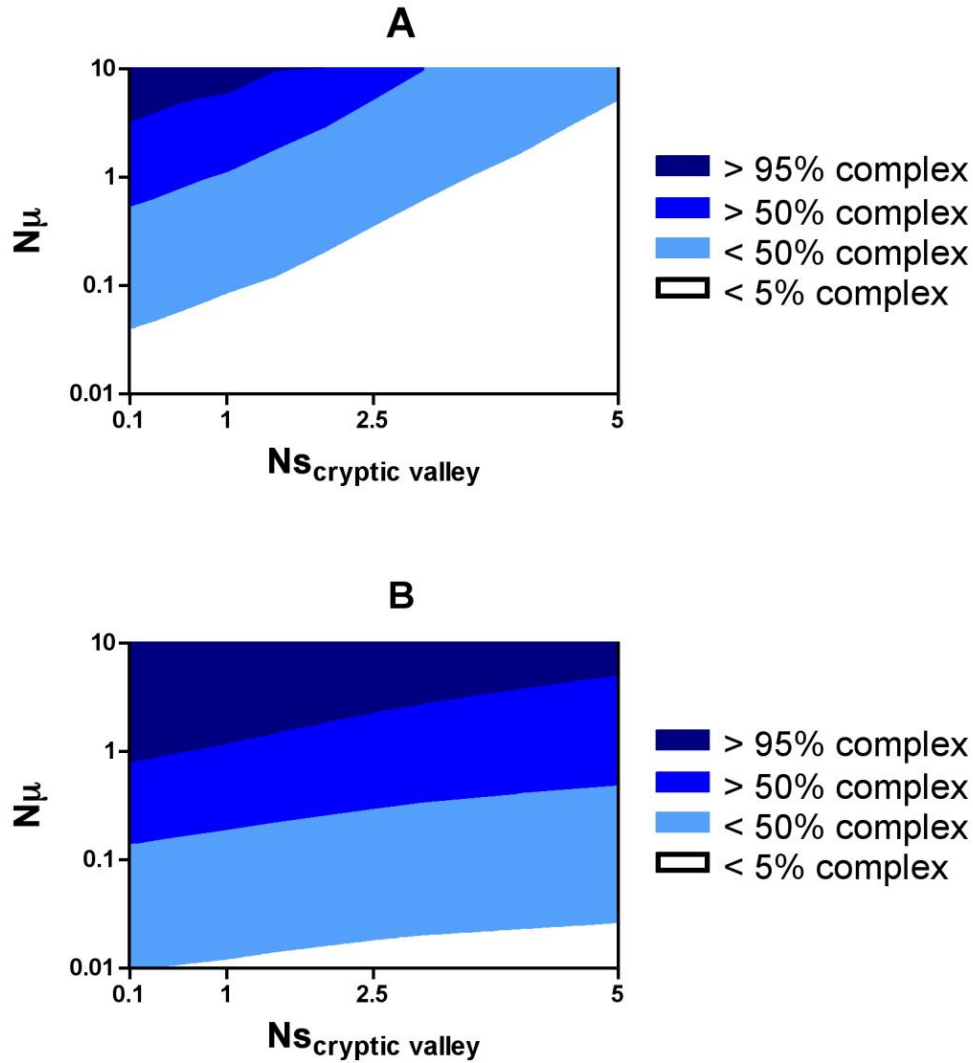

**Figure 3.** Proportion of expected fixations that are complex ($j > 1$). $Ns_{peak}= 1000$, $Ns_{valley} = 10$, $j = 1…4$. $N$ was held constant at 10,000 while $s_{cryptic\_valley}$ and $\mu$ varied. Panel A: Adaptive peak is recessive. Panel B: Adaptive peak is dominant.

We see that when mutations are significantly cryptic (a definitional criterion in our scheme), complex adaptations dominate (i.e., we find >50% of observed fixation events have $j > 1$) when the product $\mu N$ is high enough. The proportion of complex fixations is surprisingly insensitive to changes in valley depth ($s_{valley}$) (Figures S1, S2), peak height ($s_{peak}$) (Figures S3, S4), and population size $N$ (Figure S5). It is the availability of many potential adaptations, not the difficulty of valley-crossing, which drives our result.

The relevance of these results therefore depends on whether it is reasonable to assume high enough values of $\mu N$ in natural populations. Consider, for example, *Drosophila melanogaster*. Estimates based on neutral diversity assign *D. melanogaster* an $N_e \approx 10^6$ (Li and Stephan 2006). Our $\mu$ is a per-trait genomic mutation rate. If we take *D. melanogaster*'s per locus mutation rate (Haag-Liautard et al. 2007) to be on the order of $10^{-6}$, and assume an average of 10 genes per trait, our estimated per-trait rate is about $10^{-5}$. We then have $\mu N_e = 10$, far inside in the parameter range required for revelation of cryptic genetic variation to make valley-crossing a frequent source of adaptation.

With the values $N = 10^6$, $\mu = 10^{-5}$, and $s_{cryptic\_valley} = 10^{-4}$, the expected number of available 4-step potential adaptations in our model is on the order of $10^8$. While each complex adaptation remains individually unlikely, the sheer number of potential adaptations available after a capacitance event can override this individual rarity to make complexity commonplace.

## Discussion

Obviously, given the large uncertainty in the relevant parameters, we do not claim to have calculated a precise expected frequency of complex adaptations. But our assumptions would need to be wrong in a very substantial way in order to overturn our result. For example, if incrementing *j* reduces ε by many orders of magnitude, or if genomic rates of cryptic mutation are orders of magnitude below our assumed values, this might contradict our conclusions.

Some of our assumptions are quite conservative, e.g. when we consider a recessive peak. As another example of conservatism, we use the Moran model, which assumes that accidents of sampling are the primary stochastic force in molecular evolution; if linkage to other sites under selection is the primary stochastic force, complex adaptations are much more common than calculated here (Neher and Shraiman 2011).

Real adaptations are more likely clumped rather than following our assumption of a uniform distribution in genotype space. Genotypic clumps are easier to find than isolated genotypic points, since the same allele can be simultaneously favored as part of multiple possible adaptive populations. When valley crossing is rare (i.e. there are only small departures from treating $\varepsilon$ strictly as an infinitesimal), this effect is likely larger than the counteracting need to subtract double counting when multiple adaptations exist. However, as $\varepsilon$ grows larger, so does the probability that a $j$-peak is part of a clump with a second peak of complexity $<j$, reducing the effective complexity of the valley crossed. The treatment of $\varepsilon$ as an infinitesimal is useful in our simple conceptual model of a very simple genotype-fitness map, which we use to illustrate the importance of taking into account how evolutionary capacitors interact with the immense size of genotype space. Ultimately, the rate of valley crossing must of course depend on the exact shape of empirical genotype-fitness maps.

Our model assumes free recombination, raising questions as to the effects of linkage and linkage disequilibrium. Our valleys are shallow while cryptic, making valley-crossing rates with high recombination approximately equal to rates under the no-recombination limit (19, eq 2) Empirically, asexual populations of RNA enzymes that had accumulated cryptic variation also adapted more rapidly to a new substrate, via a genotype with multiple changes (Hayden et al.

2011). Thus, our findings of a strong positive effect of cryptic variation on valley crossing might not be restricted to freely recombining populations.

We consider valley crossing occurring as an adaptive response to environmental change. At the moment of environmental change, the individually deleterious alleles involved in peak crossing are at intermediate frequencies given by a quasi-stationary distribution based on the Moran model. We do not consider the possibility that one or more of the individually deleterious component alleles has already become fixed prior to the environmental change. Our work is a proof of principle that valley crossing may be common, based on one particular scenario. It is of course possible that valley crossing is also common in other scenarios not included in our model, and such a result would only add strength to our primary conclusions.

Thinking about evolution as a process often requires us to ignore our usual intuition about the threshold at which we consider improbable to become impossible. We have already become accustomed to considering the immensity of evolutionary time when we talk about adaptation. Perhaps now is the time to begin seriously considering the implications of the immensity of high-dimensional genotype space as well.

**Acknowledgments**

J.M, M.V.T, G.I.P. and K.M.P were supported by the National Institutes of Health (R01GM076041). J.M. is a Pew Scholar in the Biomedical Sciences, a Fellow at the Wissenschaftskolleg zu Berlin and is supported by NIH grant (R01GM104040) and the John Templeton Foundation. D.B.W. was supported by ERC grant 250152.  K.M.P received support from the Undergraduate Biology Research Program at the University of Arizona. G.I.P thanks Alex Lancaster for discussion.

1   **Supplementary Materials**

2   **Effects of valley depth parameter** $s_{valley}$

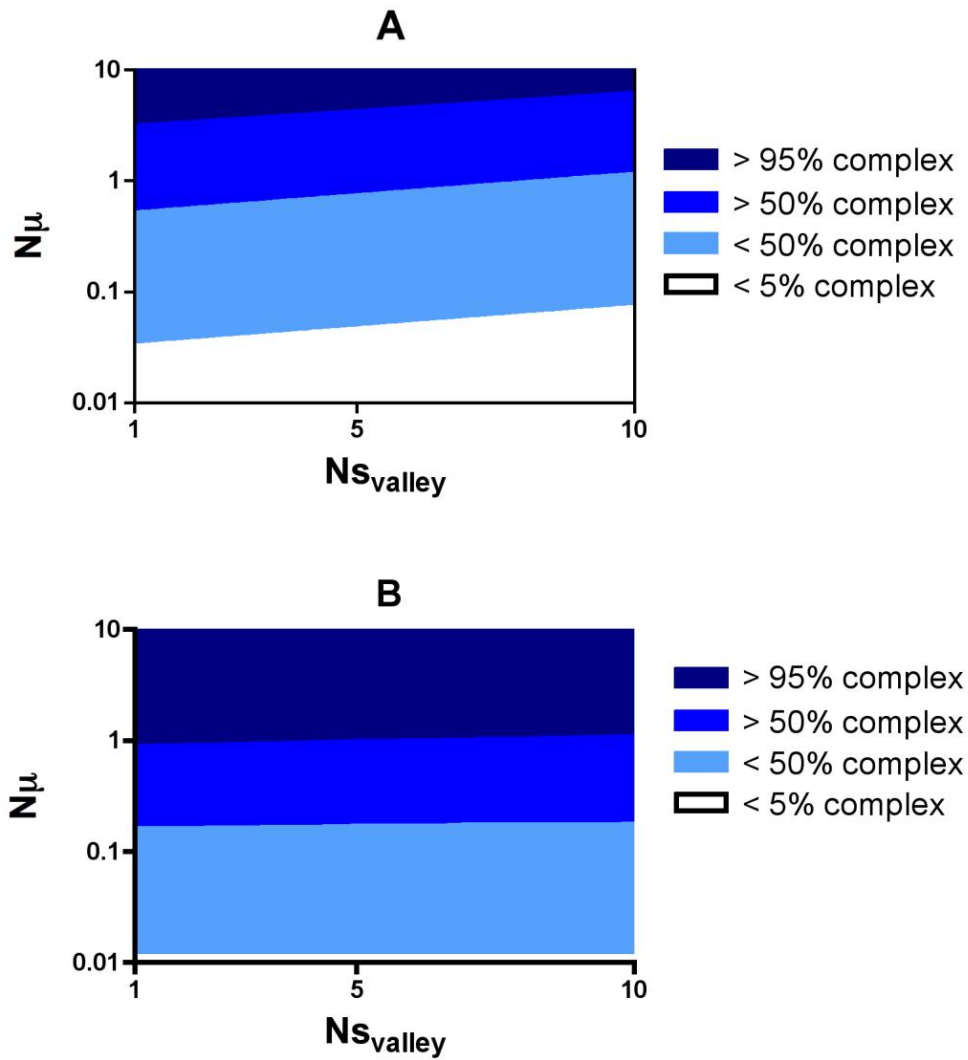

4   **Figure S1.** Percent complex fixations as a function of $s_{valley}$ and $\mu$. In both panels N=10,000,
5   $Ns_{cryptic\_valley} = 5$, and $Ns_{peak} = 1000$. We see that for recessive peaks (top panel), valley depth has
6   a moderate effect, while for dominant peaks (bottom panel), even deep valleys have only a weak
7   effect in preventing complex adaptations.

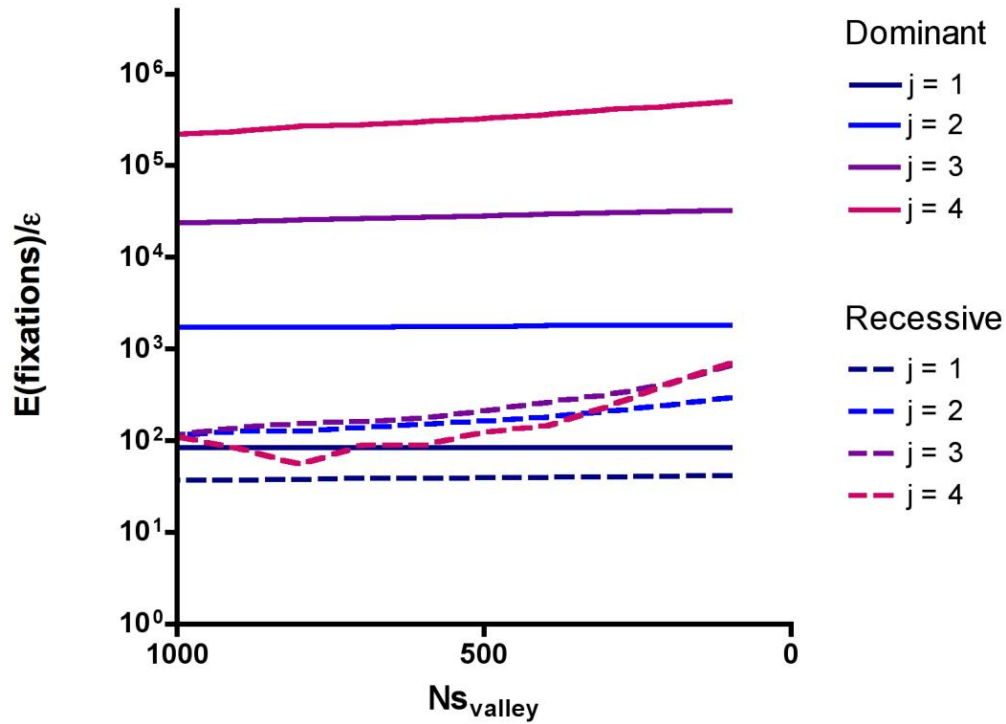

**Figure S2.** Expected numbers of fixations of $j$-sized adaptations as a function of the selection against valley genotypes, $s_{valley}$. For each value of $j$ and $s_{valley}$, we performed at least $10^5$ simulations to calculate a Monte Carlo estimate of the probability that a given $j$-sized adaptation would fix, and multiplied this by the expected number of $j$-sized adaptations that could be created by recombination of existing polymorphic alleles. Here $N = 10000$, $\mu = 0.0001$, $Ns_{peak} = 1000$ and $Ns_{cryptic\ valley} = 5$. Dashed lines: codominance is assumed both pre and post-revelation (a 'recessive' peak). Solid lines: post-revelation the adaptive peak fitness is assumed to be dominant. $Ns_{valley}$ ranges from 0.5 to 10 because in the codominant case, no fixations with $j > 2$ were observed for values $> 10$.

Percent complexity is determined primarily by the number of available j-combinations, $\frac{2\mu N t(N, s_{cryptic\_valley})^j}{j}$. This is why the level of selection on cryptic variation is the most important factor affecting the probability of crossing any given valley.

**Effects of peak height parameter,** $s_{peak}$

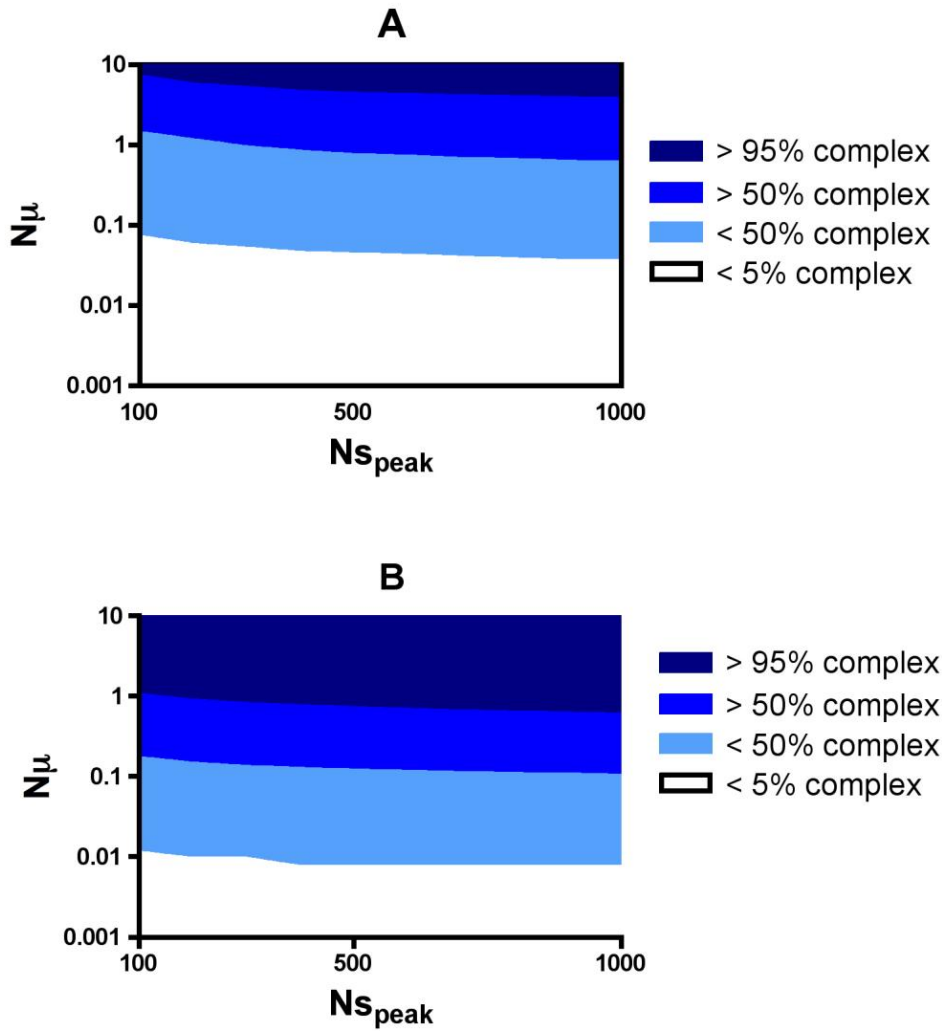

**Figure S3.** Percent complex fixations as a function of $s_{peak}$ and $\mu$. In both panels $Ns_{cryptic\ valley} = 5$, $Ns_{valley} = 10$, $N = 10\ 000$ and $\mu$ was allowed to vary. Peak height is most important for recessive peaks (top panel) than for dominant peaks (bottom panel), but in neither case does it have a strong effect on adaptive complexity.

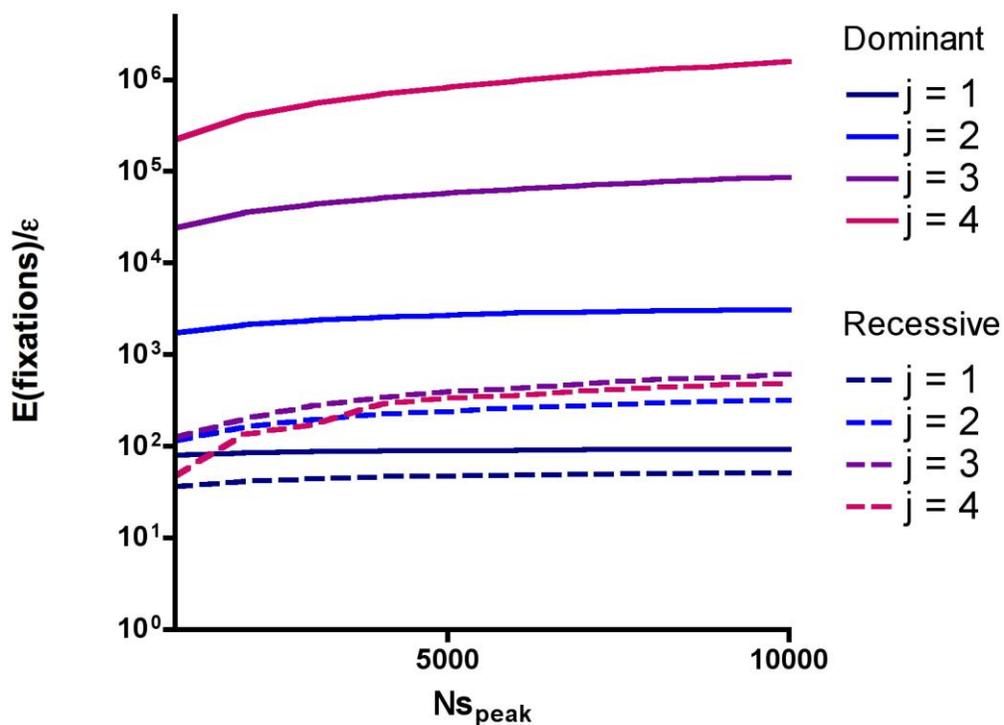

**Figure S4.** Expected numbers of fixations of *j*-sized adaptations as a function of the selection for new peak genotypes, $s_{peak}$. For each value of *j* and $s_{peak}$, we performed at least $10^5$ simulations to calculate a Monte Carlo estimate of the probability that a given *j*-sized adaptation would fix, and multiplied this by the expected number of *j*-sized adaptations that could be created by recombination of existing polymorphic alleles. Here $N=10000$, $\mu=0.0005$, $Ns_{cryptic\_valley} = 5$ and $Ns_{valley} = 10$. Dashed lines: Adaptive peak is recessive. Solid lines: Adaptive peak is dominant.

**Effects of population size *N***

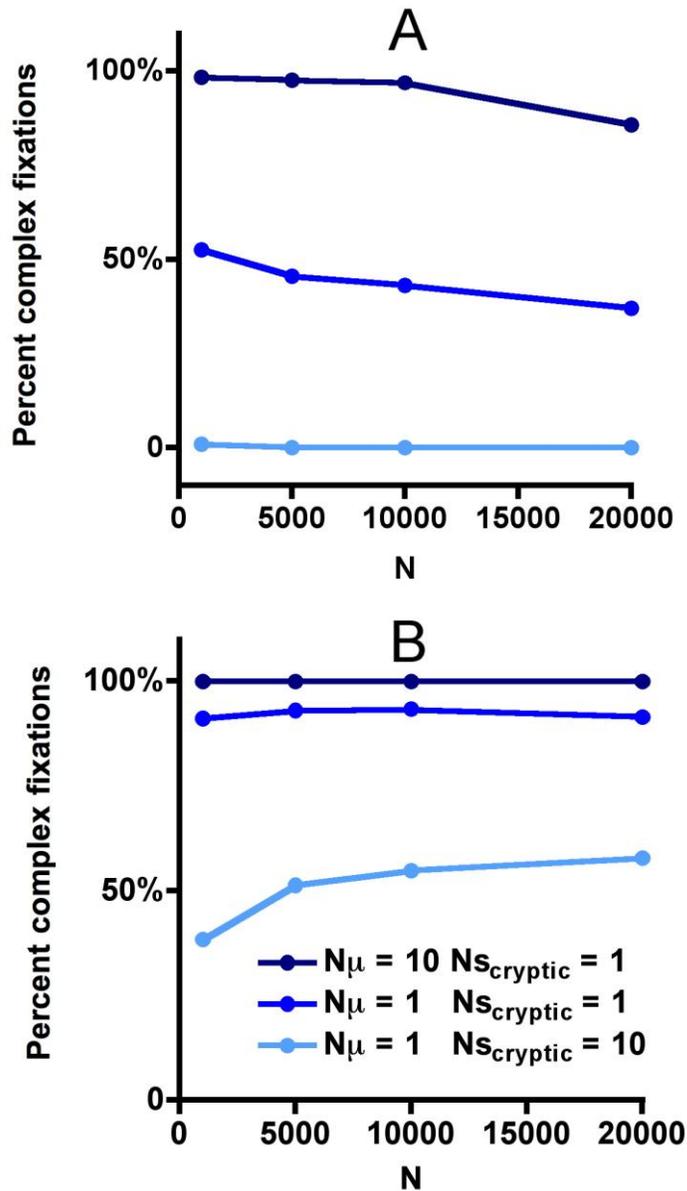

**Figure S5.** Percent complex fixations remain nearly constant across different population sizes when parameters *Nμ* and *Ns$_{cryptic\_valley}$* are kept constant. Here *Ns$_{valley}$* = 10 and *Ns$_{peak}$* = 1000. Top panel: Adaptive peak is recessive. Bottom panel: Adaptive peak is dominant.